\documentclass[12pt]{article}
\usepackage{amsmath,amssymb,bm,graphicx}
\usepackage{color} 
\usepackage{hyperref}

\newcommand{\delslash}{\not \! \partial}

\begin{document}

\begin{flushright}
\end{flushright}

\vskip 0.5 truecm

\begin{center}
{\Large{\bf Two classes of Majorana neutrinos\\ in the seesaw model }}
\end{center}
\vskip .5 truecm
\begin{center}
{\bf { Kazuo Fujikawa$^{1}$ and Anca Tureanu$^2$}}
\end{center}
\begin{center}
\vspace*{0.4cm} 
{\it {$^1$Interdisciplinary Theoretical and Mathematical Sciences Program (iTHEMS),\\
RIKEN, Wako 351-0198, Japan}
}\\
{\it {$^2$Department of Physics, University of Helsinki,\\
and Helsinki Institute of Physics,
\\P.O.Box 64, FIN-00014 Helsinki,
Finland}}
\end{center}
\makeatletter
\makeatother


\begin{abstract}
The commonly used pseudo-C symmetry $(\nu_{L})^{c}=C\overline{\nu_{L}}^{T}$ is not defined in  Lagrangian field theory. In general, there exist two classes of Majorana fermions; the first is associated with the Dirac-type fermion with  the conventional C and P symmetries, and the second is associated with the Weyl-type fermion defined by CP symmetry only and formally characterised by the pseudo-C symmetry. Taking the seesaw model as an example, it is shown that a generalized  Pauli--G\"{u}rsey (or Bogoliubov-type) canonical transformation  converts the neutrino defined by the Weyl-type fermion to the neutrino defined by the Dirac-type fermion  and thus to the conventional Majorana fermion, while preserving the canonical anti-commutation relations.   The mixing angles in the weak lepton sector are not modified by this generalized Pauli--G\"{u}rsey transformation. 

\end{abstract}


\section{Introduction}

We first summarize the conceptual complications in the case of Majorana neutrinos. One defines  the customarily used {\it pseudo-charge conjugation} transformation denoted by $\tilde{C}$ \cite{Fujikawa-Tureanu} for a left-handed Weyl fermion  
$\nu_{L}(x)$ by 
\begin{eqnarray}\label{pseudo-C}
\nu_L(x)\to\nu_{L}(x)^{\tilde{C}}=  C\overline{\nu_{L}(x)}^{T}.
\end{eqnarray}
Our notational conventions follow those in \cite{Bjorken} except for $\tilde{C}$.
Then, one may formally construct a Majorana fermion by 
\begin{eqnarray}\label{Majorana neutrino}
\psi(x) \equiv \nu_{L}(x) +C\overline{\nu_{L}}^{T}(x),
\end{eqnarray}
with $C=i\gamma^{2}\gamma^{0}$, since 
\begin{eqnarray}
\psi(x)^{\tilde{C}}=\psi(x).
\end{eqnarray}
Note the crucial property, namely, that the pseudo-C transformation \eqref{pseudo-C} changes a left-handed fermion to a right-handed fermion.
Adopting the transformation \eqref{pseudo-C} as the definition of charge conjugation leads, however, to absurd conclusions.

Suppose we construct a Lagrangian, comprising kinetic and mass terms, invariant under the pseudo-C symmetry by using \eqref{Majorana neutrino}:
\begin{eqnarray}\label{Majorana mass}
\int d^{4}x {\cal L}_{\nu} 
&=&\int d^{4}x (1/2) \{\bar{\psi} i\gamma^{\mu}\partial_{\mu}\psi(x) - \bar{\psi}(x)M_{\nu}\psi(x) \}\nonumber\\
&=&\int d^{4}x \{\overline{\nu_{L}}(x)i\gamma^{\mu}\partial_{\mu}\nu_{L}(x)
-(1/2) [\nu_{L}^{T}(x)CM_{\nu}\nu_{L}(x) + h.c.]\}
\end{eqnarray}
where $M_{\nu}$ stands for the $3\times3$ diagonalized neutrino mass matrix.   
Now, one can show that the action \eqref{Majorana mass} actually vanishes identically if the pseudo-C symmetry is invoked \cite{Fujikawa-Tureanu} as charge conjugation transformation. We re-write the Lagrangian \eqref{Majorana mass} as 
\begin{eqnarray}\label{Weinberg-like model}
\int d^{4}x {\cal L}_{\nu}
 &=&
\int d^{4}x\Big\{\overline{\nu_{L}}(x)i\gamma^{\mu}\partial_{\mu}\left(\frac{1-\gamma_{5}}{2}\right)\nu_{L}(x)\\
&-&(1/2) \left[\nu_{L}^{T}(x)CM_{\nu}\left(\frac{1-\gamma_{5}}{2}\right)\nu_{L}(x) + h.c.\right]\Big\}\nonumber.
\end{eqnarray}
In the last formula, we apply the pseudo-C operation \eqref{pseudo-C}, which is supposed to be a good symmetry, namely, $\nu_{L}(x) \rightarrow C\overline{\nu_{L}(x)}^{T}$.
Noting that $C\overline{\nu_{L}(x)}^{T}$ is right-handed, one finds that the commonly used Majorana action vanishes completely. 

This shows that the operation 
$\nu_L(x)\ \to\nu_{L}(x)^{\tilde{C}}=  C\overline{\nu_{L}(x)}^{T}$ cannot be taken as a basic definition of charge conjugation. The crucial fact is that the pseudo-C operation is not defined in  Lagrangian field theory.  
Practically, Weinberg's model \cite{Weinberg} with the form  of \eqref{Majorana mass}, for example,  is constructed with Weyl fermions in the Standard Model, and thus the Majorana fermions  are defined by CP operation since one can confirm that $\psi^{CP}(x^{0},\vec{X})=i\gamma^{0}\psi(x^{0},-\vec{x})$ for $\psi(x)$ in \eqref{Majorana neutrino}. To justify Weinberg's model, we have to resort to a notion different from the pseudo-C symmetry. We suggest the use of a canonical transformation introduced by  Pauli \cite{Pauli}, which is applicable to a  general class of vector-like theories, by regarding Weinberg's model as a limiting form of the seesaw model.

Technically, we distinguish the Weyl fermion from the chiral fermion as much as possible in the following. Only CP operation is allowed for the free Weyl fermion, while C and P are separately defined for the free chiral fermion which arises from a massive Dirac-type fermion. Thus, we define CP for the Weyl fermion
\begin{eqnarray}\label{Weyl fermion}
\nu_{L}(x^{0},\vec{x}) \rightarrow i\gamma^{0}C\overline{\nu_{L}}^{T}(x^{0},-\vec{x}),
\end{eqnarray}
but no separate C nor P. The ``Majorana'' fermion in the sense 
\begin{eqnarray}\label{Majorana neutrino2}
\psi(x) \equiv \nu_{L}(x) +C\overline{\nu_{L}}^{T}(x)
\end{eqnarray}
 is defined only for CP-symmetry as it is based on the Weyl fermion in the present case. Noting that CP conjugation is given by \eqref{Weyl fermion},
 we have the CP transformation for $\psi(x)$ given by \cite{Fujikawa-Tureanu2}
 \begin{eqnarray}\label{CP transformation}
 \psi(x^{0}, \vec{x})\rightarrow i\gamma^{0} \psi(x^{0}, -\vec{x}).
 \end{eqnarray}
However, no separate C nor P are defined for $\psi(x^{0}, \vec{x})$ in \eqref{Majorana neutrino2}.
 
 On the other hand, the chiral fermion arising from a generic Dirac-type massive fermion is defined both with the charge conjugation and parity separately by
 \begin{eqnarray}
 &&C: \ \ \ \nu_{L}(x)\rightarrow C\overline{\nu_{R}}^{T}(x)=\left(C\overline{\nu}^{T}(x)\right)_L,\nonumber\\
&& \ \ \ \ \ \ \ \ \nu_{R}(x)\rightarrow C\overline{\nu_{L}}^{T}(x)=\left(C\overline{\nu}^{T}(x)\right)_R,\nonumber\\
&&P: \ \ \ \nu_{L,R}(x^{0},\vec{x})\rightarrow i\gamma^{0}\nu_{R,L}(x^{0},-\vec{x}),
\end{eqnarray}
and thus the combined CP is formally the same as for the Weyl fermion \eqref{CP transformation}. For the convenience to deal with the Majorana fermion, we adopt the phase convention $i\gamma^{0}$ for  parity \cite{Weinberg0}.

The present paper is motivated by the fact that the most
textbooks and  papers including the influential papers on Majorana neutrinos in \cite{Bilenky} (except for \cite{Schechter2}) use the  pseudo-C symmetry, and
we believe that the problematic aspect of the pseudo-C symmetry should be appreciated more widely.

\section{Seesaw model and  generalized Pauli--G\"{u}rsey transformation}

It is instructive to analyze Majorana neutrinos in a general class of {\it seesaw models} \cite{Fukugita}, which contain  equal number of left-handed and right-handed fermions (and thus is vector-like),  
\begin{eqnarray}\label{Seesaw Lagrangian}
{\cal L} &=& \overline{\nu_{L}}(x)i\delslash\nu_{L}(x)
+\overline{\nu_{R}}(x)i\delslash\nu_{R}(x)
-\{ \overline{\nu_{L}}m_{D}\nu_{R}(x) \nonumber\\
&&+(1/2)\nu_{L}^{T}(x)C m_{L}\nu_{L}(x)
+(1/2)\nu^{T}_{R}(x)Cm_{R}\nu_{R}(x)+h.c. \}
\end{eqnarray}
where $m_{D}$ is a $3\times 3$ complex Dirac-type  mass matrix, and $m_{L}$ and $m_{R}$ are $3\times 3$ complex symmetric matrices. The anti-symmetry of the matrix $C=i\gamma^{2}\gamma^{0}$
 and Fermi statistics imply that $m_{L}$ and $m_{R}$ are symmetric. This is the Lagrangian of neutrinos with Dirac and Majorana mass terms. For $m_L=0$, it represents the classical seesaw Lagrangian of type I; namely, starting with the Dirac-type neutrinos, the very large masses $m_{R}$ are added to gauge-singlets $\nu_{R}$ and thus intuitively making $\nu_{R}$ very massive, which are consistent with the gauge structure of the SM.  In the following, we shall call the expression \eqref{Seesaw Lagrangian} as the seesaw Lagrangian for the sake of generality, but the physical analyses are performed  only for the case $m_L=0$. (In the present notation, Weinberg's model \cite{Weinberg} formally corresponds to the result after an integration over the very massive $\nu_{R}(x)$ in the type I seesaw model.) 
 
We start with the Lagrangian \eqref{Seesaw Lagrangian}, which is CP invariant but with no obvious C or (left-right) P symmetry, and write the mass term as 
\begin{eqnarray}\label{mass term}
(-2){\cal L}_{mass}=
\left(\begin{array}{cc}
            \overline{\nu_{R}}&\overline{\nu_{R}^{C}}
            \end{array}\right)
\left(\begin{array}{cc}
            m_{R}& m_{D}\\
            m_{D}^{T}&m_{L}
            \end{array}\right)
            \left(\begin{array}{c}
            \nu_{L}^{C}\\
            \nu_{L}
            \end{array}\right) +h.c.,
\end{eqnarray}
where 
\begin{eqnarray}\label{notational convention1}
\nu_{L}^{C}\equiv C\overline{\nu_{R}}^T, \ \ \ \nu_{R}^{C}\equiv C\overline{\nu_{L}}^T.  
\end{eqnarray}
Since the mass matrix appearing is complex symmetric, we can diagonalize it 
by a $6 \times 6$ unitary transformation (Autonne--Takagi factorization \cite{Autonne--Takagi}) as
\begin{eqnarray}\label{orthogonal}
            U^{T}
            \left(\begin{array}{cc}
            m_{R}& m_{D}\\
            m_{D}^{T}& m_{L}
            \end{array}\right)
            U
            =\left(\begin{array}{cc}
            M_{1}&0\\
            0&-M_{2}
            \end{array}\right)    ,        
\end{eqnarray}
where  $M_{1}$ and $M_{2}$ are $3\times 3$ real diagonal matrices (characteristic values) \footnote{We denote one of the eigenvalues as $-M_{2}$ instead of $M_{2}$ since then $M_{1},M_{2}=\sqrt{(m_{R}/2)^{2}+m_{D}^{2}}\pm m_{R}/2$, respectively,  in the case of a single flavor case with real $m_{D}$, $m_{R}$ and $m_{L}=0$ which we can solve explicitly. The mass term \eqref{mass term} is not the conventional mass term, and the Autonne--Takagi factorization has a freedom of adding an extra phase to the otherwise positive definite characteristic values. }.
We thus have
\begin{eqnarray}\label{exact-mass}
(-2){\cal L}_{mass}
&=& \left(\begin{array}{cc}
            \overline{\tilde{\nu}_{R}}&\overline{\tilde{\nu}_{R}^{C}}
            \end{array}\right)
\left(\begin{array}{cc}
            M_{1}&0\\
            0&-M_{2} 
            \end{array}\right)            
            \left(\begin{array}{c}
            \tilde{\nu}_{L}^{C}\\
            \tilde{\nu}_{L}
            \end{array}\right) +h.c.,                       
\end{eqnarray}
with
\begin{eqnarray} \label{variable-change}          
            &&\left(\begin{array}{c}
            \nu_{L}^{C}\\
            \nu_{L}
            \end{array}\right)
            = U \left(\begin{array}{c}
            \tilde{\nu}_{L}^{C}\\
            \tilde{\nu}_{L}
            \end{array}\right)
           ,\ \ \ \ 
            \left(\begin{array}{c}
            \nu_{R}\\
            \nu_{R}^{C}
            \end{array}\right)
            = U^{\star} 
            \left(\begin{array}{c}
            \tilde{\nu}_{R}\\
            \tilde{\nu}_{R}^{C}
            \end{array}\right).          
\end{eqnarray}
Hence we can write 
\begin{eqnarray}\label{exact-solution}
{\cal L}
&=&(1/2)\{\overline{\tilde{\nu}_{L}}(x)i\delslash\tilde{\nu}_{L}(x)+ \overline{\tilde{\nu}_{L}^{C}}(x)i\delslash \tilde{\nu}_{L}^{C}(x)+\overline{\tilde{\nu}_{R}}(x)i\delslash \tilde{\nu}_{R}(x)\nonumber\\
&& \ \ \ \ \ + \overline{\tilde{\nu}_{R}^{C}}(x)i\delslash \tilde{\nu}_{R}^{C}(x)\}\nonumber\\
&-&(1/2)\left(\begin{array}{cc}
            \overline{\tilde{\nu}_{R}},&\overline{\tilde{\nu}_{R}^{C}}
            \end{array}\right)
\left(\begin{array}{cc}
            M_{1}&0\\
            0&-M_{2} 
            \end{array}\right)            
            \left(\begin{array}{c}
            \tilde{\nu}_{L}^{C}\\
            \tilde{\nu}_{L}
            \end{array}\right) +h.c.\nonumber\\
 &=&(1/2)\{\overline{\psi_{+}}[i\delslash  -M_{1}]\psi_{+} +\overline{\psi_{-}}[i\delslash  - M_{2}]\psi_{-}\} ,
\end{eqnarray}
where
\begin{eqnarray}\label{pseudo-Majorana}
\psi_{+}(x)&=& \tilde{\nu}_{R}+ \tilde{\nu}_{L}^{C}=\tilde{\nu}_{R}+ C\overline{\tilde{\nu}_{R}}^{T},\nonumber\\
\psi_{-}(x)&=&\tilde{\nu}_{L}- \tilde{\nu}_{R}^{C}= \tilde{\nu}_{L}- C\overline{\tilde{\nu}_{L}}^{T},
\end{eqnarray}
with mass eigenvalues $M_{1}\neq M_{2}$.
In the present Autonne--Takagi  factorization \eqref{variable-change}, one can confirm that the crucial relations \eqref{notational convention1},
$\tilde{\nu}_{L}^{C}=C\overline{\tilde{\nu}_{R}}^{T}$ and $\tilde{\nu}_{R}^{C}=C\overline{\tilde{\nu}_{L}}^{T}$, 
 hold after the transformation \cite{KF-PG}.
These fermions \eqref{pseudo-Majorana} are formally Majorana with masses $M_{+}\neq M_{-}$ if one should use the pseudo-C symmetry, but the conventional parity is not defined.  This diagonalization is thus faithful to the starting Lagrangian \eqref{Seesaw Lagrangian}. The pseudo-C symmetry is not defined mathematically as we have already explained, and thus we should use the conventional CP symmetry to define the Majorana fermions \eqref{pseudo-Majorana} \cite{Fujikawa-Tureanu2} but without separate C or P.  The fermions in the present context are Weyl-type defined only by CP, 
\begin{eqnarray}
CP:\ \  \psi_{+}(x^{0},\vec{x})\rightarrow i\gamma^{0}\psi_{+}(x^{0}, -\vec{x}),\ \ \psi_{-}(x^{0},\vec{x})\rightarrow -i\gamma^{0}\psi_{-}(x^{0}, -\vec{x}).
\end{eqnarray}

 One may now 
	use a generalized Pauli--G\"{u}rsey \cite{Pauli} (or Bogoliubov-type) canonical transformation to change the definition of the vacuum of the Weyl-type fermion to the vacuum of the Majorana  fermion, or re-name the fermions by mixing creation and annihilation components (and thus absorbing quantum numbers C and P from the vacuum) but preserving canonical anti-commutation relations \cite{KF-PG}. 
We thus consider a further $6 \times 6$ real generalized Pauli--G\"{u}rsey transformation $O$, which has generally  the same form as the Autonne--Takagi factorization, but is orthogonal in the present specific case and thus preserves CP as will be explained later:
\begin{eqnarray} \label{Pauli--Gursey2}          
            &&\left(\begin{array}{c}
            \tilde{\nu}_{L}^{C}\\
            \tilde{\nu}_{L}
            \end{array}\right)
            = O \left(\begin{array}{c}
            N_{L}^{C}\\
            N_{L}
            \end{array}\right)
           ,\ \ \ \ 
             \left(\begin{array}{c}
            \tilde{\nu}_{R}\\
            \tilde{\nu}_{R}^{C}
            \end{array}\right)
            = O 
            \left(\begin{array}{c}
            N_{R}\\
            N_{R}^{C}
            \end{array}\right),         
\end{eqnarray}
with a specific $6 \times 6$ orthogonal transformation 
\begin{eqnarray}\label{Pauli-Gursey}
O=\frac{1}{\sqrt{2}} \left(\begin{array}{cc}
            1&1\\
            -1&1
            \end{array}\right).
\end{eqnarray}
We  then obtain \cite{KF-PG}
\begin{eqnarray}\label{transformed Majorana}
{\cal L}&=&(1/2)[\overline{N}(x)i\delslash N(x) +\overline{N^{C}}(x)i\delslash N^{C}(x) ]\nonumber\\
&&-\frac{1}{4}\{\overline{N}(x)[M_{1}+M_{2}]N(x) +\overline{N^{C}}(x)[M_{1}+M_{2} ] N^{C}(x) \} \nonumber\\
&&-\frac{1}{4}\{\overline{N}(x)[M_{1}-M_{2}]N^{C}](x) +\overline{N^{C}}(x)[M_{1}-M_{2} ] N(x) \}\nonumber\\
&=&(1/2)\{\overline{\Psi_{+}}[i\delslash  -M_{1}]\Psi_{+} +\overline{\Psi_{-}}[i\delslash  - M_{2}]\Psi_{-}\},
\end{eqnarray}
which is invariant under the standard C, P and CP defined by 
\begin{eqnarray}
&&C:\ \ N(x)\leftrightarrow N^{C}(x)=C\overline{N}^{T}(x)\nonumber\\
&&P:\ \ N(t,\vec{x})\rightarrow i\gamma^{0}N(t,-\vec{x}),\ \ N^{C}(t,\vec{x})\rightarrow i\gamma^{0}N^{C}(t,-\vec{x}),\nonumber\\
&&CP:\ \ N(x)\rightarrow i\gamma^{0}N^{C}(t,-\vec{x}),\ \ N^{C}(x)\rightarrow i\gamma^{0}N(t,-\vec{x}). 
\end{eqnarray}
From these symmetry properties, the variables $N(x)$ correspond to Dirac-type variables. We used the generalized Pauli--G\"{u}rsey transformation \eqref{Pauli-Gursey}, which is equivalent to the Bogoliubov-type transformation, but extended easily to three generations. 
In \eqref{transformed Majorana} we used the relations which are obtained from \eqref{pseudo-Majorana},  \eqref{Pauli--Gursey2} and  \eqref{Pauli-Gursey}:
\begin{eqnarray} \label{Pauli--Gursey4}          
            &&\left(\begin{array}{c}
            {\psi_{+}}_{L} \\
            {\psi_{-}}_{L}
            \end{array}\right) =\left(\begin{array}{c}
            \tilde{\nu}_{L}^{C}\\
            \tilde{\nu}_{L}
            \end{array}\right)
            = \frac{1}{\sqrt{2}} \left(\begin{array}{c}
            N_{L}^{C} + N_{L} \\
            - N_{L}^{C}+ N_{L}
            \end{array}\right) =\left(\begin{array}{c}
            {\Psi_{+}}_{L} \\
            {\Psi_{-}}_{L}
            \end{array}\right) 
           ,\nonumber\\
            &&\left(\begin{array}{c}
            {\psi_{+}}_{R} \\
           - {\psi_{-}}_{R}
            \end{array}\right) = \left(\begin{array}{c}
            \tilde{\nu}_{R}\\
            \tilde{\nu}_{R}^{C}
            \end{array}\right)
            =\frac{1}{\sqrt{2}} 
            \left(\begin{array}{c}
            N_{R} +N_{R}^{C}\\
            -N_{R} +N_{R}^{C}
            \end{array}\right)=\left(\begin{array}{c}
            {\Psi_{+}}_{R} \\
           - {\Psi_{-}}_{R}
            \end{array}\right) .        
\end{eqnarray} 
The variables $\Psi_{\pm}(x)$ satisfy the same form of Dirac equations as $\psi_{\pm}(x)$ but we assign later different physical interpretations to them (C and P properties are different although CP is the same)  and thus we use different notations\footnote{In the Pauli-Gursey canonical transformation, we have
\begin{eqnarray}
\psi_{+}&=&\tilde{\nu}_{R}+C\overline{\tilde{\nu}_{R}}^{T}\Rightarrow \frac{1}{\sqrt{2}} (N_{R} +N_{R}^{C})+\frac{1}{\sqrt{2}}(N_{L} +N_{L}^{C})=\Psi_{+}\nonumber\\
\psi_{-}&=&\tilde{\nu}_{L}-C\overline{\tilde{\nu}_{L}}^{T}\Rightarrow 
\frac{1}{\sqrt{2}}(N_{L} -N_{L}^{C})-\frac{1}{\sqrt{2}}(N_{R}^{C}-N_{R})=\Psi_{-}\nonumber.
\end{eqnarray}
Before the canonical transformation, $\tilde{\nu}_{R}$ and $C\overline{\tilde{\nu}_{R}}^{T}$, for example, are related by the pseudo-C with opposite chiralities, while these two terms are mapped to $\frac{1}{\sqrt{2}} (N_{R} +N_{R}^{C})$ and $\frac{1}{\sqrt{2}}(N_{L} +N_{L}^{C})$, respectively, which are separately invariant under the conventional C with opposite chiralities. }.
We thus have
\begin{eqnarray}\label{real Majorana}
\Psi_{\pm}(x)= \frac{1}{\sqrt{2}}[N(x) \pm N^{C}(x)]=\frac{1}{\sqrt{2}}[N(x) \pm C\overline{N}^{T}(x)],
\end{eqnarray}
with the same masses $M_{+}\neq M_{-}$ as in \eqref{pseudo-Majorana}. 
The space-time structure defined by the Dirac equations \eqref{exact-solution} and \eqref{transformed Majorana} are not modified by the present Bogoliubov-type canonical transformation, and thus only the internal freedom of C and P are re-arranged. The conventional CP symmetry is preserved in \eqref{Pauli--Gursey4} but C and P are not preserved separately. (The relations corresponding to \eqref{transformed Majorana} and \eqref{real Majorana} are obtained by a Bogoliubov-type transformation for a single flavor model \cite{Fujikawa-Tureanu}.)  

The fields $\Psi_{\pm}(x)$ in \eqref{real Majorana} are the Majorana fermions with well-defined C and P in the conventional sense. To be precise, $\Psi_{+}^{C}(x)=\Psi_{+}(x)$ and $\Psi_{-}^{C}(x)=-\Psi_{-}(x)$, and thus \eqref{transformed Majorana} is written as 
 \begin{eqnarray}\label{real Majorana1}
 {\cal L}&=&\frac{1}{2}\Psi_{+}(x)^{T}C[i\delslash - M_{+}]\Psi_{+}(x)
- \frac{1}{2}\Psi_{-}(x)^{T}C[i\delslash - M_{-}]\Psi_{-}(x). 
\end{eqnarray}
 The field $N(x)$ is a Dirac-type as being defined by a sum of two Majorana fields from \eqref{real Majorana}:
 \begin{eqnarray}\label{real Majorana2}
 N(x)&=&\frac{1}{\sqrt{2}}[ \Psi_{+}(x)+\Psi_{-}(x)],\nonumber\\
 N^{C}(x)&=&\frac{1}{\sqrt{2}}[ \Psi_{+}(x)-\Psi_{-}(x)],
 \end{eqnarray}
 but the mass eigenvalues of $\Psi_{\pm}$ are different, and thus we called $N$ and $N^C$ ``Dirac-type fermions''. The generalized Pauli--G\"{u}rsey transformation is  a canonical transformation, and thus the  anti-commutation relations are the standard ones 
 \begin{eqnarray*}\{N^{\dagger}(x^{0},\vec{x}),N(x^{0},\vec{y})\}& =&\delta^{3}(\vec{x}-\vec{y}),\cr
 \{N(x^{0},\vec{x}),N(x^{0},\vec{y})\}&=&\{N^{C}(x^{0},\vec{x}),N^{C}(x^{0},\vec{y})\}=0,
 \end{eqnarray*}
   from \eqref{real Majorana1} and \eqref{real Majorana2} for the  fermions $N(x)$ with three internal degrees of freedom.

 \section{Leptonic mixing angles and CP} 
  The identical transformation $\psi_{\pm}\rightarrow \Psi_{\pm}$ in \eqref{Pauli--Gursey4} is  important to keep the weak mixing matrix in \eqref{weak interaction} below  invariant.  One may recall the charged current weak interaction vertex in the SM
 \begin{eqnarray}
 {\cal L}_{W}=\frac{g}{\sqrt{2}}\overline{l_{L}}(x)\gamma^{\mu}W_{\mu}^{-}(x)\frac{(1-\gamma_{5})}{2}\nu_{L}(x) + h.c.,
 \end{eqnarray}
 where $l(x)$ and $\nu_{L}(x)$ are the charged lepton triplet and the neutrino  triplet (but belonging to the gauged  $SU_{L}(2)\times U(1)$) in \eqref{Seesaw Lagrangian}, respectively. 
 This interaction Lagrangian after diagonalizing the charged lepton mass and the neutrino sector by \eqref{variable-change} is given by (see also \cite{Schechter})
 \begin{eqnarray}\label{weak interaction}
 {\cal L}_{W}
 &=&\frac{g}{\sqrt{2}}\overline{l_{L}}^{\alpha}(x)\gamma^{\mu}W_{\mu}^{-}(x)[U^{\alpha k}\tilde{\nu}^{k}_{L}(x)+(U^{\prime})^{\alpha k}(\tilde{\nu}_{L}^{C})^{k}(x)] + h.c.\nonumber\\
 &=&\frac{g}{\sqrt{2}}\overline{l_{L}}^{\alpha}(x)\gamma^{\mu}W_{\mu}^{-}(x)[U^{\alpha k}{\Psi_{-}}_{L}^{k}(x)+(U^{\prime})^{\alpha k} {\Psi_{+}}_{L}^{k}(x)] + h.c.,
 \end{eqnarray}
where we used \eqref{Pauli--Gursey4} in the last line. The mixing matrices $U^{\alpha k}$ and $(U^{\prime})^{\alpha k} $ contain the products of the charged leptonic mixing matrix and the parts of the $3\times 3$ submatrices ($U_{22}$ and $U_{21}$, respectively) of $U$ in the first relation of \eqref{variable-change}. The mixing matrix $U^{\alpha k}$ is identified with the conventional Pontecorvo--Maki--Nakagawa--Sakata (PMNS) mixing matrix.   If $\frac{(1-\gamma_{5})}{2}\Psi_{-}(x)={\Psi_{-}}_{L}(x)$ is measured, $\frac{(1+\gamma_{5})}{2}\Psi_{-}(x)={\Psi_{-}}_{R}(x)$ is recovered by the parity operation, namely, the emitted free neutrinos $\Psi_{-}$ are the bona-fide Majorana particles with well-defined C and P.

	It is satisfactory that the phenomenological analyses of leptonic weak interactions is essentially the same as those in \cite{Bilenky, Schechter2} including extra CP phases up to the Bogoliubov-type re-naming \footnote{The neutrinoless double $\beta$ decay, however, vanishes for the first expession of \eqref{weak interaction} since the neutrino in the first expression is written as $(\frac{1-\gamma_{5}}{2})(\tilde{\nu}_{L}(x)-C\overline{\tilde{\nu}_{L}}^{T})$. The  neutrino field $\tilde{\nu}_{L}(x)-C\overline{\tilde{\nu}_{L}}^{T}$ and its free propagator, when identified with a Majorana neutrino defined  by the pseudo-C symmetry, vanish due to the same reasoning as \eqref{Weinberg-like model}. See \cite{Fujikawa}.}, and this may be the reason why the pseudo-C symmetry has been used widely without recognizing its mathematical inconsistency. The difference is that one now has Majorana neutrinos  in a mathematically consistent manner without referring to the pseudo-C symmetry in the second expression of \eqref{weak interaction} thanks to the generalized Pauli--G\"{u}rsey (or Bogoliubov-type) canonical transformation. The moral is that it is not sufficient to add simply a mass term to  a Weyl fermion $\nu_{L}(x)$ to convert it to a Majorana fermion. Generally, one needs to add $\nu_{R}(x)$ and re-arrange the vacuum  to define a Dirac-type fermion 
and then Majorana fermions in a consistent manner \footnote{The absence of the Majorana--Weyl fermion in $d=4$ is suggestive. 
For example,  only one of the Weyl fields in  \eqref{extra Majorana}, such as $\chi$, by itself does not determine either one of $\rho_{1}$ or $\rho_{2}$ in \eqref{ordinary transformation2} in the massless limit.}.

\section{Two classes of Majorana fermions}

Starting with the seesaw model with Weyl-type fermions \eqref{pseudo-Majorana}, we performed the generalized Pauli--G\"{u}rsey transformation  \eqref{Pauli--Gursey4} and defined the Majorana fermions \eqref{real Majorana1} and  Dirac-type fermions \eqref{real Majorana2}.
It is instructive  to look at the problem from a reversed direction, to see the appearance of two classes of Majorana fermions using the notational conventions with  $\gamma_{5}$ diagonal,
\begin{eqnarray}\label{conventional metric conventions}
\vec{\gamma}=\left(\begin{array}{cc}
            0&-i\vec{\sigma}\\
            i\vec{\sigma}&0
            \end{array}\right),
\gamma_{4}=\left(\begin{array}{cc}
            0&1\\
            1&0
            \end{array}\right),  
\gamma_{5}=\left(\begin{array}{cc}
            1&0\\
            0&-1
            \end{array}\right),  
C=\left(\begin{array}{cc}
            -\sigma_{2}&0\\
            0&\sigma_{2}
            \end{array}\right).
\end{eqnarray}            
The 4-component Dirac field in a theory of  single species is parameterized by 
\begin{eqnarray}\label{4-component Dirac}
&&\psi(x) =\left(\begin{array}{c}
            \chi\\
            \sigma_{2}\phi^{\star}
            \end{array}\right) , \ \  
\psi^{C}(x) =C\overline{\psi}^{T}(x)=\left(\begin{array}{c}
            \phi\\
            \sigma_{2}\chi^{\star}
            \end{array}\right).  
 \end{eqnarray}  
 The C transformation implies $\chi \leftrightarrow \phi$, namely, the chiral picture is
 \begin{eqnarray}
 C:\   \psi_{R}=\left(\begin{array}{c}
            \chi\\
            0
            \end{array}\right) \rightarrow                  
 C\overline{\psi_{L}}^{T}=\left(\begin{array}{c}
            \phi\\
            0
            \end{array}\right), \
 \psi_{L}=\left(\begin{array}{c}
            0\\
            \sigma_{2}\phi^{\star}
            \end{array}\right) \rightarrow                  
 C\overline{\psi_{R}}^{T}=\left(\begin{array}{c}
            0\\
            \sigma_{2}\chi^{\star}
            \end{array}\right). 
\end{eqnarray}            
The conventional Majorana fermions are defined by a rotation of $\pi/4$  
\begin{eqnarray}\label{ordinary transformation2}
\chi=\frac{1}{\sqrt{2}}(\rho_{1}+\rho_{2}),\ \
\phi=\frac{1}{\sqrt{2}}(\rho_{1}-\rho_{2})
\end{eqnarray}
and their C transformation properties are  $\rho_{1}\rightarrow \rho_{1}$ and $\rho_{2}\rightarrow -\rho_{2}$. When one defines
\begin{eqnarray}\label{conventional Majorana}
&&\psi(x)=\frac{1}{\sqrt{2}}(\psi_{M_{1}}+\psi_{M_{2}}),\ \ 
\psi_{M_{1}}=\left(\begin{array}{c}
            \rho_{1}\\
            \sigma_{2}\rho_{1}^{\star}
            \end{array}\right), \ \ 
\psi_{M_{2}}=\left(\begin{array}{c}
            \rho_{2}\\
            -\sigma_{2}\rho_{2}^{\star}
            \end{array}\right),
\end{eqnarray}                     
they satisfy the properties of the conventional Majorana fermions
\begin{eqnarray}
&&\psi_{M_{1}}=C\overline{\psi_{M_{1}}}^{T}, \ \  \psi_{M_{2}}=-C\overline{\psi_{M_{2}}}^{T},
\end{eqnarray}
consistent with C transformation properties of $\rho_{1}$ and $\rho_{2}$. Under $i\gamma_{4}$-parity 
$\psi(x) \rightarrow i\gamma_{4}
\psi(x^{4},-\vec{x})$, one confirms that
$\rho_{1}(x)\rightarrow i\sigma_{2}\rho^{\star}_{1}(x^{4},-\vec{x})$ and $\rho_{2}(x)\rightarrow -i\sigma_{2}\rho^{\star}_{2}(x^{4},-\vec{x})$. It is important that we start with a Dirac-type fermion to define the conventional Majorana fermions.

In this notation, another class of ``Majorana'' fermions are defined by 
\begin{eqnarray}\label{extra Majorana}
\psi_{+}(x)=\psi_{R}(x)+ C\overline{\psi_{R}(x)}^T=\left(\begin{array}{c}
            \chi\\
            \sigma_{2}\chi^{\star}
            \end{array}\right),\ 
\psi_{-}(x)=\psi_{L}(x)- C\overline{\psi_{L}(x)}^T=\left(\begin{array}{c}
            -\phi\\
            \sigma_{2}\phi^{\star}
            \end{array}\right),
\end{eqnarray}
which satisfy formally the Majorana conditions $\psi_{+}=C\overline{\psi_{+}}^{T}$ and $\psi_{-}=-C\overline{\psi_{-}}^{T}$, but the pseudo-C transformation $\psi_{L}\rightarrow C\overline{\psi_{L}}^{T}$, for example,  is different from what we expect for $\phi$; for this reason, besides the pseudo-C symmetry being not defined in Lagrangian field theory, these fermions are usually discarded. In some models like the seesaw model, however, this class of Majorana fermions are the primary class of fermions. These Majorana fermions are defined in terms of Weyl components and  discussed in Sections 1 
and 2; we emphasize that the fields \eqref{extra Majorana} themselves are well-defined by CP if one does  not use the pseudo-C symmetry such as $\psi_{L}\rightarrow C\overline{\psi_{L}}^{T}$. 

We now see what we have done in the seesaw model \eqref{Pauli--Gursey4}. Using a generalized Pauli--G\"{u}rsey  $canonical$ transformation, one rewrites those fermions  \eqref{extra Majorana}
in the notation of \eqref{Pauli--Gursey4} (but with the metric conventions in \eqref{4-component Dirac} and remembering \eqref{conventional Majorana})
\begin{eqnarray}
\left(\begin{array}{c}
            {\psi_{+}}_{R}\\
            {\psi_{+}}_{L}
            \end{array}\right)=\left(\begin{array}{c}
            \tilde{\nu}_{R}\\
            \tilde{\nu}_{L}^{C}
            \end{array}\right)=\left(\begin{array}{c}
            \chi\\
            \sigma_{2}\chi^{\star}
            \end{array}\right)
            &\Rightarrow&
            \left(\begin{array}{c}
           \rho_{1}\\
            \sigma_{2}\rho_{1}^{\star}
            \end{array}\right)
            =\left(\begin{array}{c}
            {\Psi_{+}}_{R}\\
            {\Psi_{+}}_{L}
            \end{array}\right),\nonumber\\
\left(\begin{array}{c}
            {-\psi_{-}}_{R}\\
            {\psi_{-}}_{L}
            \end{array}\right)=\left(\begin{array}{c}
            \tilde{\nu}_{R}^{C}\\
            \tilde{\nu}_{L}
            \end{array}\right)=\left(\begin{array}{c}
            \phi\\
            \sigma_{2}\phi^{\star}
            \end{array}\right)
            &\Rightarrow&
            \left(\begin{array}{c}
            -\rho_{2}\\
            -\sigma_{2}\rho_{2}^{\star}
            \end{array}\right)
            =\left(\begin{array}{c}
            -{\Psi_{-}}_{R}\\
            {\Psi_{-}}_{L}
            \end{array}\right), 
\end{eqnarray}
where the double arrow indicates a generalized Pauli--G\"{u}rsey (or Bogoliubov-type) canonical transformation;  the conventional CP is preserved, but C and P are not defined separately. In the ordinary  transformation \eqref{ordinary transformation2}, both C and P are preserved. The Majorana fermions \eqref{extra Majorana} are converted to \eqref{conventional Majorana} by the generalized Pauli--G\"{u}rsey canonical transformation.   If one sets $\psi_{\pm}=\Psi_{\pm}$ but with different C and P properties, the mass spectra of the systems $\chi$ and $\phi$ determine the mass spectra of the systems $\rho_{1}$ and $\rho_{2}$ after the Bogoliubov-type transformation, which in turn constrain the Dirac-type fermion, as we have demonstrated in the seesaw model. 

In conclusion,  we have shown that the seesaw model utilizes the Majorana fermions corresponding to \eqref{extra Majorana}  by clarifying how the conventional C and P are realized by the generalized Pauli--G\"{u}rsey canonical transformation\footnote{In \cite{Schechter}, results  related to ours are presented. Our formulation is, however, more transparent as to how the conventional C and P symmetries are generated from the seesaw Lagrangian which has no obvious C nor P symmetry.}.
 
\section*{Acknowledgements}
 The present work is supported in part by JSPS KAKENHI (Grant No.18K03633).

\end{document}